\documentclass[fleqn,twoside]{article}
\usepackage{espcrc2}
\usepackage{amsmath}
\usepackage{amsfonts}

\usepackage[dvips]{graphicx}

\newcommand{\del}{\partial}
\newcommand{\Z}{{\mathbb Z}}
\DeclareMathOperator{\Tr}{Tr}
\DeclareMathOperator{\re}{Re}

\newcommand{\link}{
\setlength{\unitlength}{14pt}
\begin{picture}(1,1)(0,0)
\linethickness{0.25pt}
\put(0,0){\circle*{0.15}}
\put(0,0){\vector(1,0){1}}
\put(1,0){\circle{0.14}}
\end{picture}}
\newcommand{\stapleup}{
\setlength{\unitlength}{14pt}
\begin{picture}(1,1)(0,0)
\linethickness{0.25pt}
\put(0,0){\circle*{0.15}}
\put(0,0){\vector(0,1){1}}
\put(0,1){\vector(1,0){1}}
\put(1,1){\vector(0,-1){1}}
\put(1,0){\circle{0.14}}
\end{picture}}
\newcommand{\stapledown}{
\raisebox{-14pt}{
\setlength{\unitlength}{14pt}
\begin{picture}(1,1)(0,-1)
\linethickness{0.25pt}
\put(0,0){\circle*{0.15}}
\put(0,0){\vector(0,-1){1}}
\put(0,-1){\vector(1,0){1}}
\put(1,-1){\vector(0,1){1}}
\put(1,0){\circle{0.14}}
\end{picture}}}

\title{
Dynamical fat link fermions}

\newcommand{\mybf}{\fontseries{b}\selectfont}

\author{Waseem Kamleh\address[CSSM]{Special Research Centre for the Subatomic Structure of Matter (CSSM) and Department of Physics and Mathematical Physics, University of Adelaide 5005, Australia.},
Derek B. Leinweber\addressmark[CSSM],
Anthony G. Williams\addressmark[CSSM]}

\begin{document}

\thispagestyle{empty}

\begin{abstract}

The use of APE smearing or other blocking techniques in fermion actions can provide many advantages. There are many variants of these fat link actions in lattice QCD currently, such as FLIC fermions. Frequently, fat link actions make use of the APE blocking technique in combination with a projection of the blocked links back into the special unitary group. This reunitarisation is often performed using an iterative maximisation of a gauge invariant measure. This technique is not differentiable with respect to the gauge field and thus prevents the use of standard Hybrid Monte Carlo simulation algorithms. The use of an alternative projection technique circumvents this difficulty and allows the simulation of dynamical fat link fermions with standard HMC and its variants. 

\end{abstract}

\maketitle

\section{INTRODUCTION}

The use of smeared or ``fat'' links in lattice fermion actions has been of interest for some time now\cite{Degrand}.  Fat Link Irrelevant Clover (FLIC) fermions have shown a number of promising advantages over standard actions, including improved convergence properties \cite{zanotti-hadron} and $O(a)$ improved scaling without the need for nonperturbative tuning \cite{zanotti-hadron2}. Furthermore, a reduced exceptional configuration problem has allowed efficient access to the light quark mass regime in the quenched approximation \cite{excepcfg}. However, recent progress in the field has shown that the behaviour of quenched QCD can differ from the true theory both qualitatively and quantitatively in the chiral regime \cite{zanotti03,young03}. Thus, interest is now focusing on dynamical QCD, be it (truly) unquenched, or partially quenched. 

In particular, as it is in the chiral regime where the difference from the valence approximation will be highlighted, we would like to go to light quark masses in dynamical QCD. This is an extremely computationally expensive endeavour. One might hope that the excellent behaviour at light quark mass displayed by FLIC fermions will carry over from the quenched theory to the unquenched one. This brings us to the issue of generating dynamical gauge field configurations with the fermionic determinant being that of the FLIC action.

\begin{table*}
\begin{center}
\begin{tabular}{ccc}
Sweep & Unit Circle & MaxReTr \\
\hline
  0   & {\mybf 0.866138301214314}   & {\mybf 0.866138301214314}\\
  1   & {\mybf 0.9603}13394813806   & {\mybf 0.9603}48747275940 \\
  2   & {\mybf 0.9807}35000838119   & {\mybf 0.9807}51346847750 \\
  3   & {\mybf 0.9883}84926461589   & {\mybf 0.9883}93707639555 \\
  4   & {\mybf 0.99210}3013943516   & {\mybf 0.99210}7844842705 \\
  5   & {\mybf 0.99418}2852413813   & {\mybf 0.99418}5532052157 \\ 
  6   & {\mybf 0.99545}7365275018   & {\mybf 0.99545}8835653863 \\
  7   & {\mybf 0.99629}3668622924   & {\mybf 0.99629}4454083006 \\
  8   & {\mybf 0.996878}305318083   & {\mybf 0.996878}710433084 
\end{tabular}
\end{center}
\caption{\label{tab:uzero} The mean link $u_0$ for a single configuration as a function of number of APE smearing sweeps, for the two different projection methods. The boldface indicates significant digits which match. The configuration is a dynamical gauge field with DBW2 glue and FLIC sea fermions, at $\beta = 8.0,\kappa = 0.128.$}
\end{table*}

\section{MaxReTr PROJECTION}

The standard technique for simulating dynamical fermions has for some time now been Hybrid Monte Carlo (HMC) \cite{hmc}. The HMC algorithm alternates between a global Metropolis accept/reject step and a gauge field update through the use of a hybrid Molecular Dynamics integration,
\begin{equation}
U_\mu(x,\tau + \Delta\tau) = U_\mu(x,\tau)\exp\big(i\Delta\tau P_\mu(x,\tau)\big),
\end{equation}
\begin{equation}
P_\mu(x,\tau + \Delta\tau) = P_\mu(x,\tau) - U_\mu(x,\tau)\frac{\delta S}{\delta U_\mu(x,\tau)}.
\end{equation}
Here, the $P_\mu(x)$ are the momenta canonically conjugate to the gauge field. The update of the momenta involves the derivative of the action with respect to the gauge field, a quantity which is also made use of in efficient local update algorithms. 

When attempting to use HMC with FLIC fermions, or some other fat link action for that matter, it is the calculation of $\frac{\delta S}{\delta U}$ which has proven problematic. FLIC fermions are clover-improved fermions where the irrelevant operators are constructed using APE smeared links \cite{ape-one,ape-two}. These fat links are constucted by performing several sweeps of APE smearing across the lattice. Each sweep consists of an APE blocking step,
\begin{equation}
V^{(n)}_\mu(x)[U^{(n-1)}] = (1-\alpha)\ \link + \frac{\alpha}{6} \sum_{\nu \ne \mu}\ \stapleup\ + \stapledown\ , 
\end{equation}
followed by a projection back into $SU(3)$ of the links, $U^{(n)}_\mu(x) = {\mathcal P}(V^{(n)}_\mu(x)).$ Typically, this projection is performed by iteratively maximising the following gauge invariant measure,
\begin{equation}
U^{(n)}_\mu(x)[U^{(n-1)}] = \max_{U'\in SU(3)} \re \Tr (U'(x)V^{(n)\dagger}_\mu(x)).
\end{equation}
Any fermion action which makes use of this reunitarisation method cannot be simulated using HMC, as the use of the iterative algorithm prevents one from being able to construct  $\frac{\delta S}{\delta U}$.

\section{UNIT CIRCLE PROJECTION}

Given any matrix $X$, then $X^\dagger X$ is hermitian and may be diagonalised. Then it is possible (for $\det X \ne 0$) to define a matrix \cite{liu-projection}
\begin{equation}
W = \frac{X}{\sqrt{X^\dagger X}}
\end{equation}
whose spectrum lies on the complex unit circle and is hence unitary. Furthermore, $W$ possesses the same gauge transformation properties as $X$. We can then constuct another matrix, 
\begin{equation}
W' = \frac{1}{\sqrt[3]{\det W}} W
\end{equation}
which is special unitary\footnote{The necessary cube root is absent in Ref. \cite{liu-projection}}. This technique provides an alternative method of $SU(3)$ projection (breaking the $\Z_3$ ambiguity by choosing the principal value of the cube root). Using the mean link as a measure of the smoothness of the smeared gauge field, Table \ref{tab:uzero} indicates that the two methods presented here are nearly equivalent.

However, the matrix inverse square root function can be approximated by a rational polynomial (whose poles lie on the imaginary axis),
\begin{equation}
W[X] \approx W_k[X] = d_0 X (X^\dagger X + c_0)\sum_{l=1}^{k} \frac{b_l}{X^\dagger X + c_l}. 
\end{equation}
This approximation is differentiable in a matrix sense for all $X$ for which the inverse square root can be defined. This means that we can construct $\frac{\delta S}{\delta U}$ for fermion actions which involve unit circle projection.

\section{RESULTS}

Having now written down the APE smearing prescription (with projection) in a differentiable closed form, the equations of motion necessary for the use of the HMC algorithm can be derived. Even in the simple case of APE smearing these equations are quite complex, but we will say something here about the mechanism for deriving them. Given the FLIC fermion action $S_{\rm FLIC}(U)$, which has explicit dependence upon the thin links $U$ and the fat links $U'$, it is straightforward to write down the partial matrix derivatives $\frac{\del S}{\del U}$ and $\frac{\del S}{\del U'}$. However, to write down in one step the total matrix derivative $\frac{d S}{d U}$ would be difficult, and inefficent numerically. Instead, one can use some calculus to both simplify the process and make it numerically efficient. This is done through the use of the matrix chain rule,
\begin{equation}
\frac{d S}{d U} = \frac{\del S}{\del U} + \frac{\del S}{\del U'} \star \frac{d U'}{d U}.
\end{equation}

The full details of the derivation are presented elsewhere \cite{kamleh-hmc}. We find that the numerical overhead for the equations of motion are small, and the generation of dynamical gauge fields with FLIC fermions is still dominated by the necessary conjugate gradient inversion required as part of the pseudo-fermion formulation. Results are given in Table \ref{tab:hmcresults}.

\begin{table*}[p]
\begin{center}
\begin{tabular}{ccccccccc}
$\beta$ & $\kappa$ & $S_{\rm gauge}$ & $\Delta\tau$ & $\frac{\Delta\tau_{\rm pf}}{\Delta\tau_{\rm g}}$ & $\rho_{\rm acc}$ & $u_0$ & $a$ & $m_\pi$ \\
\hline
3.6 & 0.1347 & IMP  & 0.0143 & 2 & 0.55 & 0.8226 & 0.247(9) & 0.702 \\
3.7 & 0.1340 & IMP  & 0.0147 & 2 & 0.64 & 0.8338 & 0.218(4) & 0.680 \\
3.8 & 0.1332 & IMP  & 0.0151 & 2 & 0.65 & 0.8443 & 0.180(2) & 0.738 \\
3.9 & 0.1310 & IMP  & 0.0200 & 2 & 0.66 & 0.8534 & 0.153(2) & 0.834 \\
3.9 & 0.1325 & IMP  & 0.0156 & 2 & 0.55 & 0.8540 & 0.146(2) & 0.702 \\
4.0 & 0.1301 & IMP  & 0.0200 & 2 & 0.66 & 0.8614 & 0.132(2) & 0.906 \\
4.0 & 0.1318 & IMP  & 0.0161 & 2 & 0.64 & 0.8625 & 0.121(2) & 0.799 \\
4.1 & 0.1283 & IMP  & 0.0200 & 2 & 0.75 & 0.8680 & 0.114(1) & 1.088 \\
4.1 & 0.1305 & IMP  & 0.0166 & 2 & 0.70 & 0.8685 & 0.104(1) & 0.668 \\
4.2 & 0.1246 & IMP  & 0.0200 & 2 & 0.86 & 0.8736 & 0.107(1) & 1.496 \\
4.2 & 0.1266 & IMP  & 0.0200 & 2 & 0.80 & 0.8738 & 0.097(1) & 1.346 \\
4.3 & 0.1253 & IMP  & 0.0200 & 2 & 0.83 & 0.8788 & 0.091(1) & 1.574 \\
4.4 & 0.1255 & IMP  & 0.0200 & 2 & 0.88 & 0.8836 & 0.086(1) & 1.411 \\
4.5 & 0.1253 & IMP  & 0.0200 & 2 & 0.83 & 0.8878 & 0.075(1) & 1.657 \\
4.6 & 0.1254 & IMP  & 0.0200 & 2 & 0.84 & 0.8916 & 0.072(1) & 1.617 \\
7.0 & 0.1315 & DBW2  & 0.0152 & 2 & 0.74 & 0.8344 & 0.252(6) & 0.780 \\
7.0 & 0.1345 & DBW2  & 0.0156 & 2 & 0.68 & 0.8352 & 0.233(8) & 0.673 \\
7.5 & 0.1310 & DBW2  & 0.0156 & 2 & 0.79 & 0.8516 & 0.206(3) & 0.779 \\
8.0 & 0.1305 & DBW2  & 0.0161 & 2 & 0.73 & 0.8663 & 0.168(2) & 0.764 \\
8.5 & 0.1300 & DBW2  & 0.0166 & 3 & 0.71 & 0.8774 & 0.134(1) & 0.782 \\
9.0 & 0.1224 & DBW2  & 0.0200 & 2 & 0.79 & 0.8858 & 0.137(3) & 1.412 \\
9.0 & 0.1296 & DBW2  & 0.0200 & 2 & 0.78 & 0.8865 & 0.115(1) & 0.753 \\
9.5 & 0.1228 & DBW2  & 0.0200 & 2 & 0.82 & 0.8934 & 0.109(2) & 1.576 \\
10.0 & 0.1234 & DBW2  & 0.0200 & 2 & 0.83 & 0.9000 & 0.099(2) & 1.502 \\
10.5 & 0.1236 & DBW2  & 0.0200 & 2 & 0.79 & 0.9056 & 0.093(1) & 1.567 \\
11.0 & 0.1239 & DBW2  & 0.0200 & 2 & 0.81 & 0.9110 & 0.086(1) & 1.473 \\
\end{tabular}
\end{center}
\caption{\label{tab:hmcresults} The step sizes, acceptance rate, mean link, lattice spacing and pion mass (in GeV) for different dynamical simulations, using (two flavours of) FLIC sea quarks and both Lusher-Wiesz (IMP) glue and DBW2 glue. These results are obtained from 20 $12^3 \times 24$ configurations. Simulations are done using multiple time step HMC with trajectories of unit length. The lattice spacing is set by $r_0$ via the static quark potential.}
\end{table*}

\section{CONCLUSION}

Unit circle projection can be written in closed form and represented by a rational polynomial approximation. The matrix rational polynomial is differentiable with respect to the gauge field and thus enables one to derive the equations of motions necessary for the use of HMC with FLIC fermions\footnote{ A proposal for another type of smearing scheme that is differentiable has also appeared recently\cite{stout-links}.}. This provides a significant advantage over the MaxReTr method, where the absence of a differentiable form precludes the use of HMC and hence the availability of a simulation algorithm that scales linearly with the lattice volume $V$, although there are $O(V^2)$ alternatives \cite{hasenfratz-dynamical}. This means that we now have an $O(V)$ algorithm available for the dynamical simulation of FLIC fermions. Furthermore, the method is general and can be applied to any fermion action with reuniterisation, including overlap fermions with a fat link kernel \cite{kamleh-overlap,degrand-fatover,bietenholz,kovacs}, or other types of fat link actions \cite{stephenson} that may involve alternative smearing techniques \cite{hasenfratz-hyp}.


\end{document}